\documentclass[a4paper,conference,10pt,table]{IEEEtran}
\usepackage{lipsum} 
\usepackage{cite}
\usepackage{algorithmic}
\usepackage[norelsize,boxruled]{algorithm2e}
\usepackage{url}
\usepackage{mdwmath}     
\usepackage[table]{xcolor}
\usepackage{amsfonts,amsmath,amssymb,amsxtra,amsbsy,floatflt}
\usepackage{subfig}
\usepackage{indentfirst} 
\usepackage{tabularx}
\usepackage{booktabs}
\usepackage{multirow}
\usepackage{nth}
\usepackage{comment}
\usepackage{xcolor}
\usepackage{bbold}
\usepackage{hhline}
\usepackage{siunitx}
\usepackage{booktabs}
\usepackage{graphicx}
\usepackage[bottom]{footmisc}
\usepackage{pgfplots}
\usepackage[acronyms,nonumberlist,nopostdot,nomain,nogroupskip]
{glossaries}
\usepackage{tikz}
\usepackage{pgfplots}
\usepackage{accents}

\IEEEoverridecommandlockouts
\usepackage[utf8]{inputenc}
\makeatletter
\let\oldabs\abs
\def\abs{\@ifstar{\oldabs}{\oldabs*}}

\let\oldnorm\norm
\def\norm{\@ifstar{\oldnorm}{\oldnorm*}}
\makeatother
\renewcommand{\a}{\mathbf{a}}
\renewcommand{\b}{\mathbf{b}}

\newcommand{\h}{\mathbf{h}}

\newcommand{\p}{\mathbf{p}}

\renewcommand{\v}{\mathbf{v}}
\newcommand{\w}{\mathbf{w}}
\newcommand{\x}{\mathbf{x}}

\newcommand{\0}{\mathbf{0}}
\newcommand{\1}{\mathbf{1}}

\newcommand{\G}{\mathbf{G}}
\renewcommand{\H}{\mathbf{H}}

\newcommand{\V}{\mathbf{V}}

\newcommand{\mub}{\boldsymbol{\mu}}

\newcommand{\Phib}{\mathbf{\Phi}}
\newcommand{\Real}{\mbox{$\mathbb{R}$}}
\newcommand{\Compl}{\mbox{$\mathbb{C}$}}

\newcommand{\Exp}{\mathbb{E}}

\newcommand{\herm}{\mathrm{H}}

\renewcommand{\Re}{\mathrm{Re}}
\newcommand{\tr}{\mathrm{tr}}
\newcommand{\tran}{\mathrm{T}}

\newacronym{3gpp}{3GPP}{3rd Generation Partnership Project}
\newacronym{2d}{2D}{two-dimensional}
\newacronym{adc}{ADC}{Analog to Digital Converter}
\newacronym{aoa}{AoA}{Angle of Arrival}
\newacronym{aod}{AoD}{Angle of Departure}
\newacronym{5g}{5G}{5th generation}
\newacronym{b5g}{B5G}{Beyond-5th generation}
\newacronym{4g}{4G}{4th generation}
\newacronym{aimd}{AIMD}{Additive Increase Multiplicative Decrease}
\newacronym{am}{AM}{Acknowledged Mode}
\newacronym{amc}{AMC}{Adaptive Modulation and Coding}
\newacronym{ap}{AP}{Access Point}
\newacronym{aqm}{AQM}{Active Queue Management}
\newacronym{awgn}{AGWN}{Additive White Gaussian Noise}
\newacronym{balia}{BALIA}{Balanced Link Adaptation}
\newacronym{bdp}{BDP}{Bandwidth-Delay Product}
\newacronym{bf}{BF}{Beamforming}
\newacronym{cc}{CC}{Congestion Control}
\newacronym{cu}{CU}{Central Unit}
\newacronym{ecdf}{ECDF}{Empirical Cumulative Distribution Function}
\newacronym{cn}{CN}{Core Network}
\newacronym{cqi}{CQI}{Channel Quality Information}
\newacronym{cp}{CP}{Control Plane}
\newacronym{csirs}{CSI-RS}{Channel State Information - Reference Signal}
\newacronym{dc}{DC}{Dual Connectivity}
\newacronym{dce}{DCE}{Direct Code Execution}
\newacronym{dci}{DCI}{Downlink Control Information}
\newacronym{dl}{DL}{Downlink}
\newacronym{du}{DU}{Distributed Unit}
\newacronym{dmr}{DMR}{Deadline Miss Ratio}
\newacronym{dmrs}{DMRS}{DeModulation Reference Signal}
\newacronym{e2e}{E2E}{end-to-end}
\newacronym{ecn}{ECN}{Explicit Congestion Notification}
\newacronym{edf}{EDF}{Earliest Deadline First}
\newacronym{enb}{eNB}{evolved Node Base}
\newacronym{epc}{EPC}{Evolved Packet Core}
\newacronym{es}{ES}{Edge Server}
\newacronym{fdma}{FDMA}{Frequency Division Multiple Access}
\newacronym{fdd}{FDD}{Frequency Division Duplexing}
\newacronym[firstplural=Radio Access Technologies (RATs)]{rat}{RAT}{Radio Access Technology}
\newacronym{fs}{FS}{Fast Switching}
\newacronym{ftp}{FTP}{File Transfer Protocol}
\newacronym{gnb}{gNB}{Next Generation Node Base}
\newacronym{harq}{HARQ}{Hybrid Automatic Repeat reQuest}
\newacronym{hetnet}{HetNet}{Heterogeneous Network}
\newacronym{hh}{HH}{Hard Handover}
\newacronym{hol}{HOL}{Head-of-Line}
\newacronym{ia}{IA}{Initial Access}
\newacronym{ieee}{IEEE}{Institute of Electrical and Electronics Engineers}
\newacronym{ilp}{ILP}{Integer Linear Program}
\newacronym{imt}{IMT}{International Mobile Telecommunication}
\newacronym{iot}{IoT}{Internet of Things}
\newacronym{ldpc}{LDPC}{Low-Density Parity Check}
\newacronym{los}{LOS}{Line-of-Sight}
\newacronym{lte}{LTE}{Long Term Evolution}
\newacronym{m2m}{M2M}{Machine to Machine}
\newacronym{mac}{MAC}{Medium Access Control}
\newacronym{mc}{MC}{Multi-Connectivity}
\newacronym{mcs}{MCS}{Modulation and Coding Scheme}
\newacronym{mec}{MEC}{Mobile Edge Cloud}
\newacronym{mi}{MI}{Mutual Information}
\newacronym{mimo}{MIMO}{Multiple Input, Multiple Output}
\newacronym{miso}{MISO}{Multiple Input, Single Output}
\newacronym{mmwave}{mmWave}{millimeter wave}
\newacronym{mptcp}{MPTCP}{Multipath TCP}
\newacronym{mr}{MR}{Maximum Rate}
\newacronym{mss}{MSS}{Maximum Segment Size}
\newacronym{mtd}{MTD}{Machine-Type Device}
\newacronym{mtu}{MTU}{Maximum Transmission Unit}
\newacronym{nfv}{NFV}{Network Function Virtualization}
\newacronym{nlos}{NLOS}{Non-Line-of-Sight}
\newacronym{nlosv}{NLOSv}{Vehicle Non-Line-of-Sight}
\newacronym{nr}{NR}{New Radio}
\newacronym{ofdm}{OFDM}{Orthogonal Frequency Division Multiplexing}
\newacronym{pdcch}{PDCCH}{Physical Downlonk Control Channel}
\newacronym{pdcp}{PDCP}{Packet Data Convergence Protocol}
\newacronym{pdsch}{PDSCH}{Physical Downlink Shared Channel}
\newacronym{pdu}{PDU}{Packet Data Unit}
\newacronym{pf}{PF}{Proportional Fair}
\newacronym{pgw}{PGW}{Packet Gateway}
\newacronym{phy}{PHY}{Physical}
\newacronym{pbch}{PBCH}{Physical Broadcast Channel}
\newacronym{pla}{PLA}{Planar Linear Array}
\newacronym[plural=\gls{mme}s,firstplural=Mobility Management Entities (MMEs)]{mme}{MME}{Mobility Management Entity}
\newacronym{prb}{PRB}{Physical Resource Block}
\newacronym{pss}{PSS}{Primary Synchronization Signal}
\newacronym{pscch}{PSCCH}{Physical Sidelink Control Channel}
\newacronym{pucch}{PUCCH}{Physical Uplink Control Channel}
\newacronym{pusch}{PUSCH}{Physical Uplink Shared Channel}
\newacronym{rach}{RACH}{Random Access Channel}
\newacronym{ran}{RAN}{Radio Access Network}
\newacronym{red}{RED}{Random Early Detection}
\newacronym{rf}{RF}{Radio Frequency}
\newacronym{rlc}{RLC}{Radio Link Control}
\newacronym{rlf}{RLF}{Radio Link Failure}
\newacronym{rrc}{RRC}{Radio Resource Control}
\newacronym{rrm}{RRM}{Radio Resource Management}
\newacronym{rr}{RR}{Round Robin}
\newacronym{rs}{RS}{Remote Server}
\newacronym{rsrp}{RSRP}{Reference Signal Received Power}
\newacronym{rss}{RSS}{Received Signal Strength}
\newacronym{rtt}{RTT}{Round Trip Time}
\newacronym{rw}{RW}{Receive Window}
\newacronym{rx}{RX}{Receiver}
\newacronym{sa}{SA}{standalone}
\newacronym{sack}{SACK}{Selective Acknowledgment}
\newacronym{sap}{SAP}{Service Access Point}
\newacronym{sc}{SC}{Single Carrier}
\newacronym{sch}{SCH}{Secondary Cell Handover}
\newacronym{scoot}{SCOOT}{Split Cycle Offset Optimization Technique}
\newacronym{sdma}{SDMA}{Spatial Division Multiple Access}
\newacronym{sdr}{SDR}{semi-definite relaxation}
\newacronym{sinr}{SINR}{Signal to Interference plus Noise Ratio}
\newacronym{siso}{SISO}{Single-Input-Single-Output}
\newacronym{sl}{SL}{Sidelink}
\newacronym{slnr}{SLNR}{Signal-to-Leakage-and-Noise-Ratio}
\newacronym{sm}{SM}{Saturation Mode}
\newacronym{snr}{SNR}{Signal-to-Noise-Ratio}
\newacronym{son}{SON}{Self-Organizing Network}
\newacronym{ss}{SS}{Synchronization Signal}
\newacronym{srs}{SRS}{Sounding Reference Signal}
\newacronym{sss}{SSS}{Secondary Synchronization Signal}
\newacronym{tb}{TB}{Transport Block}
\newacronym{tcp}{TCP}{Transmission Control Protocol}
\newacronym{tdd}{TDD}{Time Division Duplexing}
\newacronym{tdma}{TDMA}{Time Division Multiple Access}
\newacronym{tfl}{TfL}{Transport for London}
\newacronym{tm}{TM}{Transparent Mode}
\newacronym{trp}{TRP}{Transmitter Receiver Pair}
\newacronym{tti}{TTI}{Transmission Time Interval}
\newacronym{ttt}{TTT}{Time-to-Trigger}
\newacronym{tx}{TX}{Transmitter}
\newacronym{ue}{UE}{User Equipment}
\newacronym{ul}{UL}{Uplink}
\newacronym{uml}{UML}{Unified Modeling Language}
\newacronym{um}{UM}{Unacknowledged Mode}
\newacronym{utc}{UTC}{Urban Traffic Control}
\newacronym{vm}{VM}{Virtual Machine}
\newacronym{rsrq}{RSRQ}{Reference Signal Received Quality}
\newacronym{rssi}{RSSI}{Received Signal Strength Indicator}
\newacronym{rv}{RV}{Random Variable}
\newacronym{crs}{CRS}{Cell Reference Signal}
\newacronym{nsa}{NSA}{Non Stand Alone}
\newacronym{mrdc}{MR-DC}{Multi \gls{rat} \gls{dc}}
\newacronym{endc}{EN-DC}{E-UTRAN-\gls{nr} \gls{dc}}
\newacronym{5gc}{5GC}{5G Core}
\newacronym{si}{SI}{Study Item}
\newacronym{iab}{IAB}{Integrated Access and Backhaul}
\newacronym{wf}{WF}{Wired-first}
\newacronym{hqf}{HQF}{Highest-quality-first}
\newacronym{pa}{PA}{Position-aware}
\newacronym{mlr}{MLR}{Maximum-local-rate}
\newacronym{wbf}{WBF}{Wired Bias Function}
\newacronym{mib}{MIB}{Master Information Block}
\newacronym{sib}{SIB}{Secondary Information Block}
\newacronym{rnti}{RNTI}{Radio Network Temporary Identifier}
\newacronym{dft}{DFT}{Discrete Fourier Transform}
\newacronym{kpi}{KPI}{Key Performance Indicator}
\newacronym{ppp}{PPP}{Poisson Point Process}
\newacronym{v2v}{V2V}{Vehicle-to-Vehicle}
\newacronym{wave}{WAVE}{Wireless Access in Vehicular Environments}
\newacronym{udp}{UDP}{User Datagram Protocol}
\newacronym{upa}{UPA}{Uniform Planar Array}
\newacronym{fec}{FEC}{Forward Error Correction}
\newacronym{v2x}{V2X}{Vehicle-To-Everything}
\newacronym{psfch}{PSFCH}{Physical Sidelink Feedback Channel}
\newacronym{pssch}{PSSCH}{Physical Sidelink Shared Channel}
\newacronym{csma}{CSMA}{Carrier Sense Multiple Access}
\newacronym{v2n}{V2N}{Vehicle-to-Network}
\newacronym{wlan}{WLAN}{Wireless Local Area Network}
\newacronym{cav}{CAV}{Connected and Autonomous Vehicle}
\newacronym{v2i}{V2I}{Vehicle-to-Infrastructure}
\newacronym{d2d}{D2D}{Device-to-Device}
\newacronym{c-its}{C-ITS}{Connected Intelligent Transportation System}
\newacronym{fr2}{FR2}{Frequency Range 2}
\newacronym{bs}{BS}{Base Station}
\newacronym{sdu}{SDU}{Service Data Unit}
\newacronym{csi}{CSI}{Channel State Information}
\newacronym{scs}{SCS}{Subcarrier Spacing}
\newacronym{sumo}{SUMO}{Simulation of Urban MObility}
\newacronym{prr}{PRR}{Packet Reception Ratio}
\newacronym{edca}{EDCA}{Enhanced Distribution Channel Access}
\newacronym{sdap}{SDAP}{Service Data Adaptation Protocol}
\newacronym{thz}{THz}{terahertz}
\newacronym{scm}{SCM}{Spatial Channel Model}
\newacronym{vr}{VR}{Virtual Reality}
\newacronym{6g}{6G}{6th generation}
\newacronym{qos}{QoS}{Quality of Service}
\newacronym{uav}{UAV}{unmanned Aerial Vehicles}
\newacronym{bap}{BAP}{Backhaul Adaptation Protocol}
\newacronym{ns3}{ns-3}{Network Simulator 3}
\newacronym{rl}{RL}{Reinforcement Learning}
\newacronym{ris}{RIS}{Reconfigurable Intelligent Surface}
\newacronym{ula}{ULA}{Uniform Linear Array}
\newtheorem{problem}{Problem}

\newcommand{\rmB}{\scriptscriptstyle \mathrm{B}}
\newcommand{\rmR}{\scriptscriptstyle \mathrm{R}}
\newcommand{\rmBL}{{\scriptscriptstyle \mathrm{BL}}}
\newcommand{\AF}{\mathrm{AF}}

\title{A Leakage-based Method for Mitigation of Faulty Reconfigurable Intelligent Surfaces}
\author{
     \IEEEauthorblockN{Nairy Moghadas Gholian\IEEEauthorrefmark{1},~\IEEEmembership{Student Member,~IEEE,} Marco Rossanese\IEEEauthorrefmark{2},~\IEEEmembership{Student Member,~IEEE,}\\Placido Mursia\IEEEauthorrefmark{2},~\IEEEmembership{Member,~IEEE,}
     Andres Garcia-Saavedra\IEEEauthorrefmark{2},~\IEEEmembership{Member,~IEEE,}, Arash Asadi\IEEEauthorrefmark{1},~\IEEEmembership{Senior Member,~IEEE},\\Vincenzo Sciancalepore\IEEEauthorrefmark{2},~\IEEEmembership{Senior Member,~IEEE}, and Xavier Costa-Pérez\IEEEauthorrefmark{2},~\IEEEmembership{Senior Member,~IEEE}}
     \IEEEauthorblockA{\IEEEauthorrefmark{1}Technical University of Darmstadt (TUDa), Darmstadt, Germany\\ngholian@wise.tu-darmstadt.de, aasadi@wise.tu-darmstadt.de\\\IEEEauthorrefmark{2}NEC Laboratories Europe GmbH, Heidelberg, Germany\\\{name.surname\}@neclab.eu
     }
      	\thanks{This work was supported by EU FP for Research and Innovation
Horizon 2020 under Grant Agreements No. 861222 (MINTS), No. 101017011 (RISE-6G), and No. 101017109 (DAEMON) projects.}
 }
\date{\today}

\begin{document}
\maketitle

\begin{abstract}
\glspl{ris} are expected to be massively deployed in future beyond-5th generation wireless networks, thanks to their ability to programmatically alter the propagation environment, inherent low-cost and low-maintenance nature. Indeed, they are envisioned to be implemented on the facades of buildings or on moving objects. However, such an innovative characteristic may potentially turn into an involuntary negative behavior that needs to be addressed: an undesired signal scattering. 
In particular, \gls{ris} elements may be prone to experience failures due to lack of proper maintenance or external environmental factors. While the resulting \gls{snr} at the intended \gls{ue} may not be significantly degraded, we demonstrate the potential risks in terms of unwanted spreading of the transmit signal to non-intended \glspl{ue}. In this regard, we consider the problem of mitigating such undesired effect% caused by faulty \gls{ris} elements
by proposing two simple yet effective algorithms, which are based on maximizing the \gls{slnr} over a predefined \gls{2d} area and are applicable in the case of perfect channel-state-information (CSI) and partial CSI, respectively. Numerical and full-wave simulations demonstrate the added gains compared to leakage-unaware and reference schemes.
\end{abstract}


\glsresetall
%%%%%%%%%%%%%%%%%%%%%%%%%%%%%%%%%%%%%%%%%%%%%%%%%%%%%%%%%%%%%%%%%%
\section{Introduction}\label{sec:intro}
%%%%%%%%%%%%%%%%%%%%%%%%%%%%%%%%%%%%%%%%%%%%%%%%%%%%%%%%%%%%%%%%%%

%There is a steady expansion of the mobile data, which is resulting in a shortage in the available bandwidth in the sub-6 GHz frequencies \cite{Rap2013}. 
%Networks on high frequencies need to be deployed in order to keep up with the rapidly growing number of connected devices and numerous types of communication, such as satellite communications, \gls{uav}, \gls{v2x}, etc., where there is a pivotal need for low latency feedback to maintain a robust network. The \glspl{mmwave} frequencies can effectively solve the bandwidth shortage challenge. Yet, \glspl{mmwave} face their own set of challenges, comparable to that of other technology. These challenges include substantial propagation loss and penetration loss, particularly in cases of obstacles. 
The proliferation of \gls{ris}-assisted wireless networks has sparked a surge of research into the design and optimization of these systems, focusing on achieving congruent goals of improved spectral efficiency, enhanced coverage, and reduced energy consumption. Indeed, the \gls{ris} technology is considered the enabler of the \emph{Smart Radio Environment} for future wireless networks~\cite{Renzo2019SmartRE}, owing to its ability to manipulate the signal from the transmitter to the receiver~\cite{Wu2019}. In this regard, extensive efforts have been devoted towards demonstrating the feasibility of these surfaces through proof-of-concepts~\cite{Ross22}, and realistic modelling~\cite{Mursia23}. 
%\glspl{ris} have caught a notable amount of attention over the past few years. Many research works have investigated various use cases of \glspl{ris} for improving the wireless networks' performance. Some works looked into the energy efficiency of \gls{ris}-assisted wireless network by 
%%%%%%%%%%%%%%%%%%%%%%%%%%%%%%%%%%%%%%%%%%%%%%%%%%%%%%%%%%%%%%%%%%
% \textbf{1-2 sentences to relate the work to the scope of MINTS (V2X)}
%%%%%%%%%%%%%%%%%%%%%%%%%%%%%%%%%%%%%%%%%%%%%%%%%%%%%%%%%%%%%%%%%%
%

However, similarly to any other piece of hardware, \glspl{ris} can fail. This may be due to a variety of causes, including aging, external phenomena (installation on moving objects~\cite{Mursia2021_RIFE}), environmental factors, or disaster situations~\cite{Kisseleff2021}. Moreover, due to the large number of elements in \glspl{ris}, the maintenance and replacement of faulty \glspl{ris} may be difficult and costly to carry out, especially  when installed on building facades.
%They may be prone to failure for various causes, including in disaster situations \cite{Kisseleff2021}, due to aging, or external phenomena (installed on moving objects \cite{Mursia2021_RIFE}), and environmental factors.
%For example in disaster situations \cite{Kisseleff2021}, or due to external phenomena (installed on moving objects \cite{Mursia2021_RIFE}). Moreover, when installed on buildings, the maintenance may be difficult to be carried out.
% Moreover, it will increase operational costs. 

The problem of faulty antennas has been extensively studied in the field of phased arrays, which share several similarities with \glspl{ris}. Identification of faulty elements is mainly achieved via compressed sensing~\cite{xiong2019compressed, eltayeb2018compressive} in which external probes collect radiation patterns in the near-field and identify the abnormalities in post-processing. In~\cite{keizer2007element}, the authors compute the excitation weights for a faulty planar array in order to recover the original array pattern, while other mitigation techniques imply the use of specific hardware and/or active devices~\cite{Kuan1993}. However, in the case of \glspl{ris}, such methods cannot be applied since purely passive \gls{ris} elements cannot amplify impinging signals, and they cannot prevent failing elements from reflecting signals like active transceivers can.

Conversely, the literature investigating faulty \glspl{ris} is relatively scarce and focuses mainly on the \gls{ris} diagnostics point of view, i.e., detecting which elements are faulty and their associated state in terms of signal attenuation and phase shift. The authors of~\cite{Sun2021} propose methods that, even with limited or absent \gls{csi}, can pinpoint the antenna elements that are behaving irregularly. In~\cite{ozturk2023ris}, two different diagnosis strategies are developed to determine faulty elements and consequently perform user localization. Similarly, the authors of~\cite{Ma2021} propose an algorithm for faulty element diagnosis via compressed sensing. In~\cite{Li2020}, the authors formulate an equivalent channel estimation problem and detect the state of faulty \gls{ris} elements by specific pilot transmissions.

%{\color{red}\lipsum[1]}
In this paper, we consider a \gls{ris}-aided wireless network where a subset of the available \gls{ris} elements experiences failures. Moreover, \emph{unlike existing works that only deal with the detection of faulty \gls{ris} elements via advanced channel estimation techniques~\cite{Sun2021, ozturk2023ris, Li2020, Ma2021}, we study the problem of mitigating the associated negative effects on the system performance.} Specifically, we point out that, even in cases when faulty RIS elements do not degrade the \gls{snr} at the intended \gls{ue} significantly, the information signal is spread along unwanted directions causing a sharp increase in \emph{signal leakage} to unintended \glspl{ue} (causing interference) or potential eavesdroppers.

To address this problem, we develop a mathematical approach for optimizing the \gls{ris} configuration that maximizes the \gls{slnr} by exploiting only geometrical information on the \gls{2d} area wherein the intended \gls{ue} is located, thus effectively overcoming the negative impact of the faulty elements. To the best of our knowledge, our approach is the first to consider the \gls{slnr} as optimization metric in \gls{ris}-aided networks, and is applicable both for the case of perfect \gls{csi}, i.e., when  the position and state of the faulty elements is known, and in the case of partial \gls{csi}, i.e., when the state of the faulty elements is unknown. We provide numerical results demonstrating remarkable benefits with our proposed approaches of up to $35$\% and $20$\% improvement in \gls{slnr} under perfect and partial \gls{csi}, respectively, as compared to \emph{agnostic} schemes that ignore the presence of faulty elements, and \emph{naive} approaches that \emph{simply} aim at maximizing \gls{snr}, thus ignoring the signal leakage. Such gains are achieved at a small cost of less than $4$\% in terms of \gls{snr}.

\textbf{Notation}. Matrices and vectors are denoted in uppercase and lowercase bold font, respectively. $(\cdot)^{\tran}$, $(\cdot)^{\herm}$, and $\tr(\cdot)$ stand for transposition, Hermitian transposition, and trace of a square matrix, respectively. $\|\cdot\|$ denotes the Euclidean norm, while $|\cdot |$ and $\angle \cdot$ denote the absolute value and phase of a complex number, respectively. Lastly, $j=\sqrt{-1}$ is the imaginary number, and $\mathrm{diag}(\x)$ is a square matrix whose diagonal is equal to $\x$ and all other elements are zero.%$\0_{N \times M}$ the all-zero $N\times M$ matrix, Lastly, 

%%%%%%%%%%%%%%%%%%%%%%%%%%%%%%%%%%%%%%%%%%%%%%%%%%%%%%%%%%%%%%%%%%
\section{System model}\label{sec:System_model}
%%%%%%%%%%%%%%%%%%%%%%%%%%%%%%%%%%%%%%%%%%%%%%%%%%%%%%%%%%%%%%%%%%

\begin{figure}[t]
        \center
        \includegraphics[width=0.7\columnwidth, height=0.6\columnwidth]{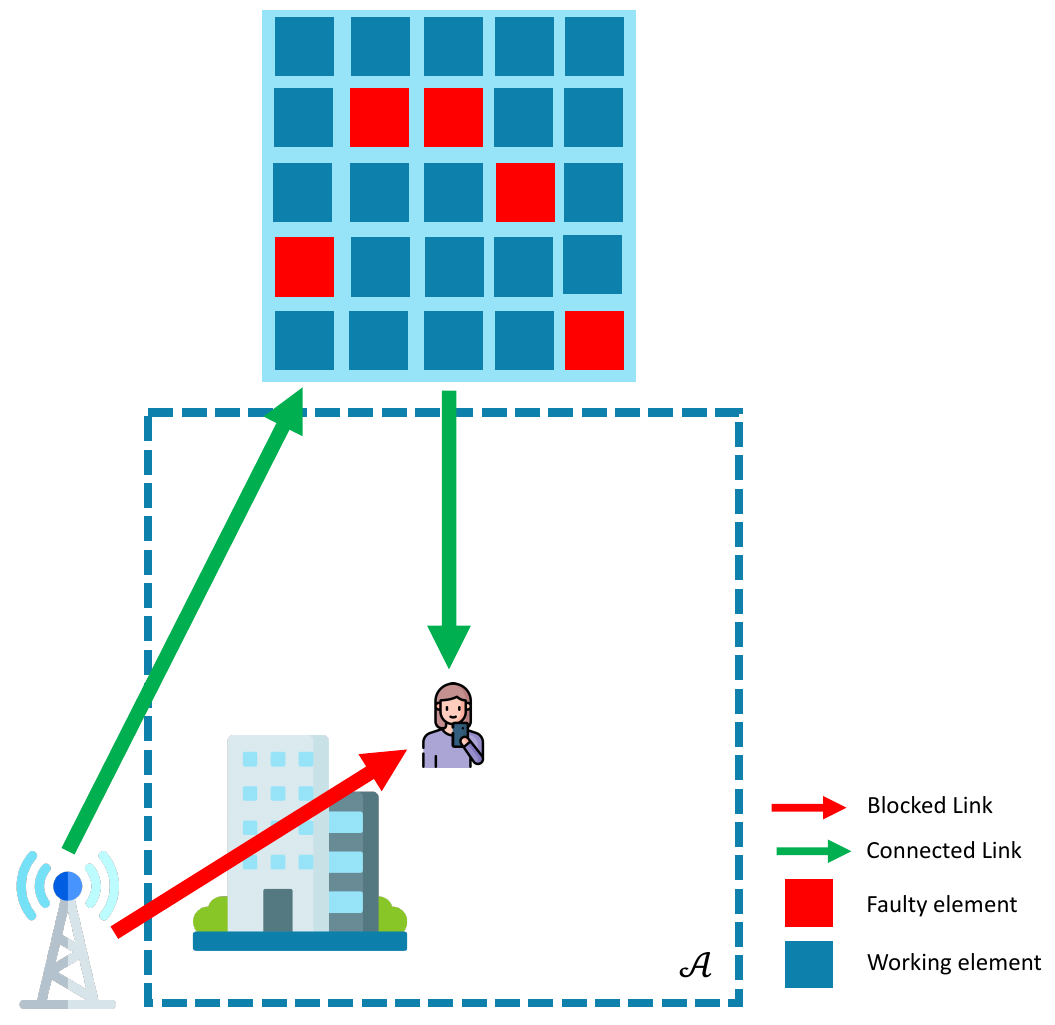}
        \caption{\small System model.}
        \label{fig:Fig1}
\end{figure}

We consider a downlink \gls{miso} wireless network,\footnote{Note that, in order to simplify the presentation, we focus on a \gls{miso} system. However, the extension to a \gls{mimo} setting is readily obtained by simply implementing receive combining.} wherein an \gls{ap} equipped with $M$ antennas, a \gls{ris} equipped with $N$ antennas in \gls{los} with the \gls{ap}, and an intended \gls{ue} are deployed, as depicted in Fig.~\ref{fig:Fig1}. Furthermore, we assume that the direct link between the \gls{ap} and the \gls{ue} is blocked, i.e., in \gls{nlos}. Let $\h_k \in\Compl^{N \times 1}$ and $\G \in\Compl^{N \times M}$ denote the channel from the \gls{ris} to \gls{ue} $k$ and from the \gls{ap} to the \gls{ris}, respectively. Hence, the receive signal at the intended \gls{ue} $k$ is given by
\begin{align}
    y_k = \h_k^\herm \Phib \G \w_k s + n \in \Compl,\label{eq:yk}
\end{align}
where $\Phib = \mathrm{diag}[e^{j\theta_1},\ldots,e^{j\theta_N}]\in \Compl^{N\times N}$ denotes the matrix of \gls{ris} phase shifts, $\w_k = \sqrt{P} \frac{\G^\herm \Phib^{\herm}\h_k}{\|\G^\herm \Phib^{\herm}\h_k\| } \in \Compl^{M\times 1}$ is the maximum-ratio transmission (MRT) precoder that maximizes the \gls{snr} for the effective channel between the \gls{ap} and the \gls{ue} through the \gls{ris}, with $P$ the power budget, $s\in \Compl$ is the transmit symbol, with $\Exp[|s|^2]=1$, and $n\in \Compl$ is the noise coefficient distributed as $\mathcal{CN}(0,\sigma_n^2)$. For simplicity, we rewrite the received signal in~\eqref{eq:yk} as
\begin{align}
    y_k & = \v^\herm \bar{\H}_k \w_k s + n,\label{eq:yk2}\\
    & = \sqrt{P} \|\v^\herm \bar{\H}_k\| s + n
\end{align}
where $\v = \mathrm{diag}(\Phib^\herm)\in\Compl^{N\times 1}$, $\bar{\H}_k = \mathrm{diag}(\h_k^\herm)\G\in\Compl^{N\times M}$, and we have used the expression of the precoder.
% According to the closed-form expression for MRT precoder, we can write $\w_k$ as:
% \begin{align}
%     \w_k = \sqrt{P} \frac{\G^\herm \Phib^{\herm}\h_k}{\|\G^\herm \Phib^{\herm}\h_k\| }
% \end{align}
%%%%%%%%%%%%%%%%%%%%%%%%%%%%%%%%%%%%%%%%%%%%%%%%%%%%%%%%%%%%%%%%%%
\subsection{Baseline model}\label{subsec:BL}
%%%%%%%%%%%%%%%%%%%%%%%%%%%%%%%%%%%%%%%%%%%%%%%%%%%%%%%%%%%%%%%%%%
We define the \gls{snr} at the intended \gls{ue} location as
\begin{align}
    \mathrm{\gls{snr}} = \frac{P\,\| \v^\herm \bar{\H}_k\|^2}{\sigma_n^2}.
\end{align}
Hence, the objective of the baseline scheme is formalized as
\begin{align}
    \begin{array}{cl}\label{eq:P_BL}
        \displaystyle \max_{\v} &  \mathrm{\gls{snr}}\\
         \mathrm{s.t.} & |v_{n}| = 1, \quad n=1,\ldots,N, 
    \end{array}
\end{align}
whose closed-form solution is given by
\begin{align}
    \v_{BL} = e^{j\angle\bar{\h}_{k,1} } \in \Compl^{N\times 1},
\end{align}
where the subscript $BL$ stands for \emph{baseline}, $\bar{\h}_{k,1}$ is the eigenvector associated to the only non-zero eigenvalue of $\bar{\H}_{k}$, which is a rank-$1$ matrix due to the \gls{los} channel between the \gls{ap} and the \gls{ris}.

%%%%%%%%%%%%%%%%%%%%%%%%%%%%%%%%%%%%%%%%%%%%%%%%%%%%%%%%%%%%%%%%%%
\subsection{Faulty RIS model}\label{subsec:Faulty_RIS}
%%%%%%%%%%%%%%%%%%%%%%%%%%%%%%%%%%%%%%%%%%%%%%%%%%%%%%%%%%%%%%%%%%

We assume that a total of $B\leq N$ \gls{ris} elements are faulty, such that $\v = [\v_{\rmR}^\tran\,\,\v_{\rmB}^\tran]^\tran$ and $\bar{\H}_k = [\bar{\H}_{\rmR,k}^\tran\,\,\bar{\H}_{\rmB,k}^\tran]^\tran$, where each $v_{\rmB,b}, \,\,b=1,\ldots,B$ is a \gls{rv} with $|v_{\rmB,b}|\leq 1$, $B\leq N$, and $\angle v_{\rmB,b}\in [0,2\pi)$, $\bar{\H}_{\rmR,k}\in \Compl^{N-B\times M}$ and $\bar{\H}_{\rmB,k}\in \Compl^{B\times M}$. In this regard, the subvector $\v_{\rmR}\in\Compl^{N-B\times 1}$ represents the tunable \gls{ris} elements that are fully functioning, whereas $\v_{\rmB}\in\Compl^{B\times 1}$ denotes the uncontrollable faulty elements. We thus rewrite the receive signal in~\eqref{eq:yk2} as
\begin{align}
    y_k & = \begin{bmatrix}\v_{\rmR}^\herm & \v_{\rmB}^\herm \end{bmatrix} \begin{bmatrix}
   \bar{\H}_{\rmR,k} \\ \bar{\H}_{\rmB,k}\end{bmatrix} \w_k s + n\\
   & = (\v_{\rmR}^\herm\bar{\H}_{\rmR,k} + {\h}^\herm_{\rmB,k})\w_k s + n,\label{eq:yk_b}
\end{align}
where we have defined 
\begin{align}
    {\h}^\herm_{\rmB,k}& =\v_{\rmB}^\herm\bar{\H}_{\rmB,k}\in \Compl^{1\times M},
\end{align}
which is treated as a fixed channel component, since it is uncontrollable by the \gls{ris}. Note that with this simple modelling, the received signal in \eqref{eq:yk_b} is in the form of an equivalent \gls{los} \gls{miso} \gls{ris}-aided network~\cite{Mursia2021}. The \gls{snr} is thus re-written as
\begin{align}
    \mathrm{\gls{snr}} = \frac{P\,\|\v_{\rmR}^\herm\bar{\H}_{\rmR,k} + {\h}^\herm_{\rmB,k} \|^2}{\sigma_n^2}.\label{eq:SNR_2}
\end{align}

%%%%%%%%%%%%%%%%%%%%%%%%%%%%%%%%%%%%%%%%%%%%%%%%%%%%%%%%%%%%%%%%%%
\section{Problem formulation and solution}\label{sec:problem}
%%%%%%%%%%%%%%%%%%%%%%%%%%%%%%%%%%%%%%%%%%%%%%%%%%%%%%%%%%%%%%%%%%

In this section, we analyze potential mitigation strategies under the assumption of perfect or partial \gls{csi} information at the \gls{ap}, i.e., the exact values of $\v_{\rmB}$, or only the indexes of the faulty elements, respectively.

%%%%%%%%%%%%%%%%%%%%%%%%%%%%%%%%%%%%%%%%%%%%%%%%%%%%%%%%%%%%%%%%%%
\subsection{Maximum SNR}
%%%%%%%%%%%%%%%%%%%%%%%%%%%%%%%%%%%%%%%%%%%%%%%%%%%%%%%%%%%%%%%%%%

As a first simple mitigation technique, we analyze the problem of maximizing the \gls{snr} at the intended \gls{ue}. Given the equivalent model in~\eqref{eq:yk_b} and~\eqref{eq:SNR_2}, 
\begin{comment}
it is easy to find the optimal \gls{ris} configuration that maximizes the \gls{snr}, which is given by
\begin{align}
    \v_{\rmR, Naive} = e^{j(\angle\bar{\h}_{\rmR,k} - \angle{h}_{\rmB,k})}.\label{eq:vR_SNR}
\end{align}    
\end{comment}
let us define 
\begin{align}\label{eq:V_R}
    \V_{\rmR} = \begin{bmatrix}\v_{\rmR}\\1\end{bmatrix}\begin{bmatrix}\v_{\rmR}^\herm & 1\end{bmatrix}\in\Compl^{\bar{N}+1\times \Bar{N}+1},    
b\end{align}
where $\bar{N}=N-B$ is the number of functioning \gls{ris} elements. Hence, the RIS configuration that maximizes the SNR under RIS element failures is given by
\begin{problem}\label{eq:P0}
\begin{align}
    \begin{array}{cl}
         \displaystyle \max_{\V_{\rmR}\succeq \0} &  \displaystyle \tr(\V_{\rmR}\widetilde{\H}_k ) \\
        \mathrm{s.t.} & \mathrm{diag}(\V_{\rmR}) = \1, \quad \mathrm{rank}(\V_{\rmR}) = 1,
    \end{array}
\end{align}
\end{problem}
where we have defined
\begin{align}\label{eq:Hb_k}
    \widetilde{\H}_k = \begin{bmatrix}\bar{\H}_{\rmR,k} \\ {\h}^\herm_{\rmB,k}\end{bmatrix}\begin{bmatrix}\bar{\H}^\herm_{\rmR,k} & {\h}_{\rmB,k}\end{bmatrix}\in\Compl^{\bar{N}+1\times \Bar{N}+1}.
\end{align}
Problem~\ref{eq:P0} can be solved by \gls{sdr}, i.e., ignoring the rank-one constraint and then approximating the \gls{ris} configuration $\v_{\rmR}$ via randomization techniques~\cite{Luo2010}.
Note that the \gls{ris} configuration obtained as a result of the aforementioned procedure aims at aligning the signal bouncing off the \emph{functioning} \gls{ris} elements with the (random) channel component that is in the direction of the \gls{ue} and originated from the superposition of the signal reflected by the \emph{faulty} \gls{ris} elements. We denote this method as \emph{naive} RIS optimization.

%%%%%%%%%%%%%%%%%%%%%%%%%%%%%%%%%%%%%%%%%%%%%%%%%%%%%%%%%%%%%%%%%%
\subsection{Maximum SLNR}
%%%%%%%%%%%%%%%%%%%%%%%%%%%%%%%%%%%%%%%%%%%%%%%%%%%%%%%%%%%%%%%%%%

In this case, our aim is not only to improve the \gls{snr} at the intended receiver by compensating for the effect of the faulty \gls{ris} elements, but to jointly minimize potential leakage to non-intended \glspl{ue} or eavesdroppers in the proximity of the \gls{ue} as well. Moreover, our proposed approach exploits geometrical information only, without the need of acquiring costly \gls{csi} of potentially non-intended \glspl{ue}. In this regard, we assume that all \glspl{ue} are located within a three-dimensional area $\mathcal{A}$. Let $\p\in \mathcal{A}\subset \Real^3$ be the location of a potential non-intended \gls{ue}, such that the \emph{leakage} over such area is given by
\begin{align}
    \mathrm{L}(\mathcal{A}) = \int_{\mathcal{A}} \|\v_{\rmR}^\herm\bar{\H}_{\rmR}(\p) + {\h}^\herm_{\rmB}(\p)\|^2 \mathrm{d}\p,
\end{align}
where $\bar{\H}_{\rmR}(\p)\in\Compl^{N\times M}$ is the equivalent channel from the \gls{ap} to the point $\p$, through the functioning \gls{ris} elements, and ${\h}^\herm_{\rmB}(\p)\in \Compl^{1\times M}$ is the equivalent direct and uncontrollable channel between the \gls{ap} and position $\p$, through the faulty \gls{ris} elements. Therefore, we define the \gls{slnr} as
\begin{align}\label{eq:SLNR_integral}
    \mathrm{SLNR} = \frac{ \|(\v_{\rmR}^\herm\bar{\H}_{\rmR,k} + {\h}^\herm_{\rmB,k})\|^2}{\mathrm{L}(\mathcal{A})  + \sigma_n^2/P}.
\end{align}
Given the complex structure of \eqref{eq:SLNR_integral}, we discretize the area $\mathcal{A}$ into $T$ \emph{test points} such that the \gls{slnr} is approximated as
\begin{align}\label{eq:SLNR_approx}
    \mathrm{SLNR} \approx \frac{ \|(\v_{\rmR}^\herm\bar{\H}_{\rmR,k} + {\h}^\herm_{\rmB,k}) \|^2}{\sum_{t=1,t\neq k}^T \|(\v_{\rmR}^\herm\bar{\H}_{\rmR,t} + {\h}^\herm_{\rmB,t})\|^2 + \sigma_n^2/P},
\end{align}
which is a sampled version of~\eqref{eq:SLNR_integral}. However, the objective function in~\eqref{eq:SLNR_approx} might be maximized by focusing on minimizing the denominator, i.e., the leakage, at the cost of a reduction in the numerator, i.e., the useful signal power. Hence, to avoid obtaining a trivial solution, we add a minimum \gls{snr} requirement, which is necessary to decode the signal, and formulate the following optimization problem
\begin{problem}\label{eq:P1}
\begin{align}
    \begin{array}{cl}
        \displaystyle \max_{\v_{\rmR}} &  \mathrm{SLNR}\\
         \mathrm{s.t.} & \|(\v_{\rmR}^\herm\bar{\H}_{\rmR,k} + {\h}^\herm_{\rmB,k})\|^2\geq \gamma\\
         & |v_{\rmR,n}| = 1, \quad n=1,\ldots,\bar{N}, 
    \end{array}
\end{align}    
\end{problem}
where the system parameter $\gamma$ regulates the trade-off between leakage reduction and useful signal power.  

By using the definition in~\eqref{eq:V_R}, we have that
\begin{align}
    \begin{array}{cl}\label{eq:P2}
    \|(\v_{\rmR}^\herm\bar{\H}_{\rmR,k} + {\h}^\herm_{\rmB,k})\|^2 & = \bigg\| \begin{bmatrix}\v_{\rmR}^\herm & 1\end{bmatrix}\begin{bmatrix}\bar{\H}_{\rmR,k} \\ {\h}^\herm_{\rmB,k}\end{bmatrix}\bigg\|^2 \\
         & = \tr(\widetilde{\H}_k\V_{\rmR}),
    \end{array}
\end{align}
where $\widetilde{\H}_k$ is defined in~\eqref{eq:Hb_k}.

By making use of the bisection method, we can rewrite Problem~\ref{eq:P1} as~\cite{Boyd04}
\begin{problem}\label{eq:P3}
\begin{align}
    \begin{array}{cl}
   \displaystyle \max_{\beta\geq 0,\,\V_{\rmR}\succeq \0}  &  \beta\\
         \mathrm{s.t.} &  \tr(\widetilde{\H}_k\V_{\rmR})\geq \beta \,(\sum_{t=1,t\neq k}^T\tr(\widetilde{\H}_t\V_{\rmR})\!+\!\sigma_n^2/P)\\
         & \tr(\widetilde{\H}_k\V_{\rmR})\geq \gamma\\
         & \mathrm{diag}(\V_{\rmR}) = \1,\,\,\mathrm{rank(\V_{\rmR})}=1.
    \end{array}
\end{align}    
\end{problem}
Problem~\ref{eq:P3} is solved via \gls{sdr}, i.e., by ignoring the non-convex rank constraint and then extracting a rank-$1$ solution via randomization techniques~\cite{Luo2010}. The resulting algorithm is formalized in Algorithm~\ref{alg:A1}. 
%{\color{red}Second possibility: use Fractional Programming on \eqref{eq:P1}-- Update: tested in Matlab, gives similar results both in terms of performance and simulation time.}

%%%%%%%%%%%%%%%%%%%%%%%%%%%%%%%%%%%%%%%%%%%%%%%%%%%%%%%%%%%%%%%%%%
\subsection{Partial CSI case: robust solution}
In the following, we describe a faulty RIS mitigation strategy, which relaxes the assumption of perfect CSI and considers only knowledge of the position of the faulty elements, and not their actual phase-shifting value. In this regard, we consider the maximization of the average SLNR, where the average is taken over the random realizations of the phase shift and amplitude attenuation applied at the faulty elements.
However, since the SLNR in~\eqref{eq:SLNR_approx} is a fractional function, for the sake of simplicity, and given the independence among the numerator and denominator, we employ Jensen's inequality and optimize a lower bound on the expected SLNR as
\begin{align}
    \Exp[\mathrm{SLNR}]\! \geq \! \frac{\Exp[ \|(\v_{\rmR}^\herm\bar{\H}_{\rmR,k} + {\h}^\herm_{\rmB,k}) \|^2]}{ \sum_{t=1,t \neq k}^T \Exp[\|(\v_{\rmR}^\herm\bar{\H}_{\rmR,t} \!+ \!{\h}^\herm_{\rmB,t})\|^2] \!+ \!\sigma_n^2/P}.
\end{align}
As explained in Section~\ref{subsec:Faulty_RIS}, we model the effect caused by the faulty elements as $v_{\rmB,i} = \delta_i e^{j\phi_i}$ where $\delta_i\sim\mathcal{U}[0,1]$ and $\phi_i\sim\mathcal{U}[0,2\pi)$. Hence, we have that
\begin{align}
    \Exp[ \|(\v_{\rmR}^\herm\bar{\H}_{\rmR,k} + {\h}^\herm_{\rmB,k}) \|^2] & =  \|\v_{\rmR}^\herm\bar{\H}_{\rmR,k}\|^2 + \Exp[\|\v_{\rmB}^\herm\bar{\H}_{\rmB,k}\|^2]\nonumber\\
    & + 2\,\Re\{\v_{\rmR}^\herm\bar{\H}_{\rmR,k}\bar{\H}_{\rmB,k}^\herm\Exp[\v_{\rmB}]\}.
\end{align}
We now examine $\Exp[\v_{\rmB}]$ as 
\begin{align}
    \Exp[v_{\rmB,i}]& = \Exp[\delta_i e^{j\phi_i}] \\  \nonumber
    & = \Exp[\delta_i](\Exp[\cos(\phi_i)]+j\Exp[\sin(\phi_i)])\\ \nonumber
    & = 0 \quad \forall i,
\end{align}
which leads to 
\begin{align}
    \Exp[\|\v_{\rmB}^\herm\bar{\H}_{\rmB,k}\|^2] & = \sum_i  \Exp[|\v_{\rmB}^\herm \bar{\h}_{\rmB,k,i}|^2]\\
    & = \sum_i \Exp\bigg[ \bigg\rvert\sum_j v_{\rmB,j}^* \bar{H}_{\rmB,k,i,j}\bigg\rvert^2\bigg],\label{eq:E_SLNR}
\end{align}
where $\bar{\h}_{\rmB,k,i}$ is the $i$-th column of $\bar{\H}_{\rmB,k}$. The expression in~\eqref{eq:E_SLNR} leads to $\Exp[\|\v_{\rmB}^\herm\bar{\H}_{\rmB,k}\|^2]= \frac{1}{3}\|\bar{\H}_{\rmB,k}\|_{\mathrm{F}}^2$ given the independence among the states of the faulty RIS elements and the fact that $\Exp[\delta_i^2] = 1/3$. Similarly, we obtain $\Exp[\|\v_{\rmB}^\herm\bar{\H}_{\rmB,t}\|^2]= \frac{1}{3}\|\bar{\H}_{\rmB,t}\|_{\mathrm{F}}^2, \quad \forall t$. The lower bound on the expected SLNR is thus rewritten as

\begin{align}
    \Exp[\mathrm{SLNR}]\!\geq\! \frac{ \|\v_{\rmR}^\herm\bar{\H}_{\rmR,k}\|^2+\frac{1}{3}\|\bar{\H}_{\rmB,k}\|_{\mathrm{F}}^2}{\sum_{t=1,t \neq k}^T \|\v_{\rmR}^\herm\bar{\H}_{\rmR,t}\|^2\!+\!\frac{1}{3}\|\bar{\H}_{\rmB,t}\|_{\mathrm{F}}^2 \!+ \!\sigma_n^2/P}.
\end{align}

Therefore, we obtain a robust mitigation solution for faulty RISs by applying Algorithm~\ref{alg:A1} with $\h_{\rmB,k} = \h_{\rmB,t} = \0, \quad \forall t$ and by adding the constant offsets $\frac{1}{3}\|\bar{\H}_{\rmB,k}\|_{\mathrm{F}}^2$ and $\sum_{t=1,t\neq k}^T \frac{1}{3}\|\bar{\H}_{\rmB,t}\|_{\mathrm{F}}^2 $ to the LHS and the RHS of the first constrain in Problem~\eqref{eq:P3}, respectively.
\setlength{\textfloatsep}{0.1cm}
\setlength{\floatsep}{0.1cm}
 \begin{algorithm}[t]
    \caption{\small Iterative algorithm for Problem~\ref{eq:P3}}
    \label{alg:A1}
    % \footnotesize
    \DontPrintSemicolon
    Initialize $h^{(0)}$, $l^{(0)}$, $\delta$, and $\gamma$ to feasible values\;
        \Repeat(\hfill\emph{Bisection loop (over $i$)}){$|h^{(i)}-l^{(i)}|/|h^{(i)}| < \delta$}{
            Set $\beta = \frac{h^{(i)}+l^{(i)}}{2}$\;
            \eIf{Problem~\ref{eq:P3} admits a a feasible solution $\V_{\rmR}^\star$ by using SDR}{
            Set $l^{(i+1)} = \beta$ and $h^{(i+1)} = h^{(i)}$\;
            }
            {
            Set $h^{(i+1)} = \beta$ and $l^{(i+1)} = l^{(i)}$\;
            }
        }
        Extract $\v_{\rmR}^\star$ from $\V_{\rmR}^\star$ via Gaussian randomization\;
  \end{algorithm}

%%%%%%%%%%%%%%%%%%%%%%%%%%%%%%%%%%%%%%%%%%%%%%%%%%%%%%%%%%%%%%%%%%
\section{Numerical results and discussion}\label{sec:results}
%%%%%%%%%%%%%%%%%%%%%%%%%%%%%%%%%%%%%%%%%%%%%%%%%%%%%%%%%%%%%%%%%%

%{\color{red}\lipsum[1-2]}
In this section, we assess the performance of our proposed frameworks targeting the maximization of the instantaneous and average \gls{slnr}, against the reference schemes, namely the leakage-unaware \emph{baseline} and the \emph{naive} approach targeting the maximization of the \gls{snr}, by evaluating both the obtained \gls{slnr} and \gls{snr} in realistic settings.

%We consider a Rician channel model for the \gls{ris}-\gls{ue} link, and a \gls{los} channel for the \gls{ap}-\gls{ris} link, which are explained in section~\ref{sec:channelmodel}. Lastly, the simulation setup, including the leakage assessment scenario, has been described in section~\ref{sec:scenario}.

%The simulation results are averaged over $1000$ realizations of the channels and the randomly generated values and indices of the broken elements and other system settings, including the coordinates of randomly generated points in the leakage regions.
%providing plausible region settings for the potential leakage points that can appear in a wireless network. 

%Later, we define the region from which we gathered our numerical results.

%%%%%%%%%%%%%%%%%%%%%%%%%%%%%%%%%%%%%%%%%%%%%%%%%%%%%%%%%%%%%%%%%%
\subsection{Channel model}\label{sec:channelmodel}
%%%%%%%%%%%%%%%%%%%%%%%%%%%%%%%%%%%%%%%%%%%%%%%%%%%%%%%%%%%%%%%%%%

%We consider a \gls{miso} network where an \gls{ap} serves as the \gls{tx} and a \gls{ue} serves as the \gls{rx}. 
We consider a Rician channel model for the \gls{ris}-\gls{ue} link, and a \gls{los} channel for the \gls{ap}-\gls{ris} link
The channel between the \gls{ris} and the \gls{ue} is given by~\cite{Mursia2021}
\begin{equation}\label {eq1}
\h_k \triangleq \sqrt {\frac {K_{\mathrm {R}}}{1+K_{\mathrm{R}}}} \, \h_k ^{\mathrm{LoS}} + \sqrt {\frac {1}{1+K_{ \mathrm {R}}}}\,\h_k^{ \mathrm {NLoS}} \in { \mathbb {C}} ^{N\times 1},
\end{equation}
%where $\mathbf{g}^{\mathrm{LoS}}$ and $\mathbf {g}^{ \mathrm {NLoS}}$ are the direct link and the link through scatterers from AP to RIS, respectively.
where $K_{\mathrm{R}}$ is the Rician factor. Moreover, the \gls{los} link from the \gls{ris} to the \gls{ue} is defined as
\begin{equation}\label{eq2}
\h_k^{ \mathrm {LoS}}\triangleq \sqrt {\gamma _{g}}\, \b{(\psi_{k})} \in {\mathbb{C}}^{N\times 1},
\end{equation}
where the distance-dependent pathloss is represented as $\gamma _{g}=\zeta_0/(d_2)^{\eta_r}$, with $\zeta_0$, $d_{2}$, and ${\eta_r}$ being the free-space loss factor at a reference distance of one meter~\cite{balanis2016antenna}, the distance between the \gls{ris} and the \gls{ue}, and the corresponding pathloss exponent, respectively. ${\mathbf b(\psi_{k})}$ is the array steering vector for the \gls{aod} $\psi_{k}$ defined as  
%the matrix that models the response of the \gls{pla} for the steering angle $\psi_{A}$, which is define as:
\begin{align} \label{eq3}
\mathbf {b}(\psi_{k}) =
& [1, e^{j\frac{2\pi}{\lambda}d\cos(\psi_k)},  \ldots, e^{j(N-1)\frac{2\pi}{\lambda}d\cos(\psi_k)}]^\tran\!\in\! {\mathbb{C}}^{N\times 1},
\end{align}
where $\lambda$ represents the signal wavelength, and $d=\lambda/2$ the inter-element spacing. Note that the \gls{los} \gls{ap}-\gls{ris} link $\G$ is obtained in a similar manner as
\begin{align}
    \G\triangleq \sqrt {\gamma _{i}}\, \b{(\psi_{A})}\a{(\psi_{D})}^\herm \in {\mathbb{C}}^{N\times M},
\end{align}
where $\psi_{A}$, $\psi_{D}$ are the \gls{aoa} and \gls{aod}, respectively, while $\a{(\psi_{D})}\in\Compl^{M\times 1}$ is the steering vector at the \gls{ap}, and $\gamma _{i} = \zeta_0 /(d_{1})^{\eta_i}$ is the distance-dependent pathloss for the \gls{ap}-\gls{ris} link, with $d_1$ and $\eta_i$ the distance between the \gls{ap} and the \gls{ris}, and its associated pathloss exponent, respectively. Lastly, the \gls{nlos} link from the \gls{ris} to the \gls{ue} is defined as 
\begin{equation}\label{eq4}
\h_k^ {\mathrm {NLoS}}\triangleq \sqrt {\frac {\gamma _{g}}{P_{K}}}\sum _{p=1}^{P_{K}}\mathbf {G}^{(w)}_{p}\circ \b(\psi _{A,p}) \in {\mathbb{C}}^{N\times 1},
\end{equation}
where $P_{K}$ represents the total number of scattering paths, $\G^{(w)}_p$ is the small scale fading coefficient of the $p$-th path with $\mathrm {vec}(\mathbf {G}^{(w)}_{p})\sim \mathcal {CN}(\mathbf {0}, \mathbf {I}_{N})$, $\circ$ denotes the Hadamard product, and $\psi _{A,p}$ is the \gls{aod} of the $p$-th path. 

%%%%%%%%%%%%%%%%%%%%%%%%%%%%%%%%%%%%%%%%%%%%%%%%%%
\subsection{Simulation setup}\label{sec:scenario}
%%%%%%%%%%%%%%%%%%%%%%%%%%%%%%%%%%%%%%%%%%%%%%%%%%

We consider a wireless network wherein an \gls{ap} equipped with a $16$-antenna \gls{ula} is assisted by a \gls{ris} with $N=N_x N_y$ elements where $N_x=10$ elements are placed on the x-axis, and $N_y=10$ elements are placed on the y-axis. We assume that the \gls{ue}, \gls{ris}, and the \gls{ap} are located at the three-dimensional coordinates of $\p_{\scriptscriptstyle  \mathrm{UE}}=(16, 16, 0)$, $\p_{\scriptscriptstyle  \mathrm{RIS}}=(10, 34, 10)$, and $\p_{\scriptscriptstyle  \mathrm{AP}}=(0, 0, 10)$, respectively. The \gls{ap} is assumed to be transmitting at the carrier frequency of $30$~GHz with a power of $P = 12$~dBm per subcarrier while the noise power is assumed to be $\sigma_{n}^2=-80$~dBm. For the \gls{ris}-\gls{ue} link, we account for $P_K =10$ different scattering paths, and we set the Rician factor to $K_{R}=10$ dB and the pathloss exponent as $\eta_r=2$. The threshold \gls{snr} value $\gamma$ in Algorithm~\ref{alg:A1} is set by dividing the \gls{snr} of the naive approach by $1.5$. The intended \gls{ue} is located at the center of a target area $\mathcal{A}$ of dimension $30\times30$~m. We sample such area in a set of $T=125$ uniformly scattered \emph{leakage points}. Moreover, we assume that the \gls{ris} elements fail according to a uniform distribution, unless otherwise stated, and we average our results over $10^3$ independent realizations of the faulty \gls{ris} elements position and state, as well as the location of the leakage points. All relevant simulation parameters are summarized in Table~\ref{tab:params}. 
\begin{comment}
    \begin{table}[h!]
\caption{\small Simulation parameters.}
\label{tab:params}
\centering
\resizebox{\linewidth}{!}{%
%\renewcommand{\arraystretch}{1.0}
\begin{tabular}{|c|c|c|c|}
\hline
\cellcolor[HTML]{EFEFEF} \textbf{Parameter} & \textbf{Value} &\cellcolor[HTML]{EFEFEF} \textbf{Parameter} & \textbf{Value} \\
\hline
 \cellcolor[HTML]{EFEFEF}  $N$  & $100$ & \cellcolor[HTML]{EFEFEF} $\lambda$ &  $10$ mm\\
\hline 
 \cellcolor[HTML]{EFEFEF} $\sigma_n^2$ & $-80$ dBm &\cellcolor[HTML]{EFEFEF} $P$ & $12$ dBm  \\
\hline 
\cellcolor[HTML]{EFEFEF}  $K_{\mathrm{R}}$  & $10$ dB & \cellcolor[HTML]{EFEFEF} $P_K$ &  $10$ \\
\hline 
\cellcolor[HTML]{EFEFEF}  $\eta_{i}$  & $2$ &\cellcolor[HTML]{EFEFEF} $\eta _{r}$ &  $2$ \\
\hline
\cellcolor[HTML]{EFEFEF}  $\p_{\scriptscriptstyle  \mathrm{AP}}$  & $(0, 0, 10)$ &\cellcolor[HTML]{EFEFEF} $\p_{\scriptscriptstyle  \mathrm{RIS}}$ &  $(10, 34, 10)$\\
\hline
\cellcolor[HTML]{EFEFEF}  $\p_{\scriptscriptstyle  \mathrm{UE}}$  & $(16, 16, 0)$ &\cellcolor[HTML]{EFEFEF} $T$ &  $125$\\
\hline
\end{tabular}
}
\renewcommand{\arraystretch}{1}
\end{table}
\end{comment}

\begin{table}[h!]
\caption{\small Simulation parameters.}
\label{tab:params}
\centering
\resizebox{\linewidth}{!}{%
\renewcommand{\arraystretch}{0.9}
\begin{tabular}{|c|c|c|c|c|c|}
\hline
\cellcolor[HTML]{EFEFEF} \textbf{Parameter} & \textbf{Value} &\cellcolor[HTML]{EFEFEF} \textbf{Parameter} & \textbf{Value} & \cellcolor[HTML]{EFEFEF}\textbf{Parameter} & \textbf{Value} \\
\hline
 \cellcolor[HTML]{EFEFEF}  $N$  & $100$ & \cellcolor[HTML]{EFEFEF} $\lambda$ &  $10$ mm &\cellcolor[HTML]{EFEFEF} $\sigma_n^2$ & $-80$ dBm\\
\hline 
\cellcolor[HTML]{EFEFEF} $P$ & $12$ dBm  & \cellcolor[HTML]{EFEFEF}  $K_{\mathrm{R}}$  & $10$ dB & \cellcolor[HTML]{EFEFEF} $P_K$ &  $10$\\
\hline 
\cellcolor[HTML]{EFEFEF}  $\eta_{i}$  & $2$ &\cellcolor[HTML]{EFEFEF} $\eta _{r}$ &  $2$ & \cellcolor[HTML]{EFEFEF}  $\p_{\scriptscriptstyle  \mathrm{AP}}$ & $(0, 0, 10)$\\
\hline
\cellcolor[HTML]{EFEFEF} $\p_{\scriptscriptstyle  \mathrm{RIS}}$ &  $(10, 34, 10)$ & \cellcolor[HTML]{EFEFEF}  $\p_{\scriptscriptstyle  \mathrm{UE}}$  & $(16, 16, 0)$ &\cellcolor[HTML]{EFEFEF} $T$ &  $125$\\
\hline
\end{tabular}
}
\renewcommand{\arraystretch}{1}
\end{table}

\begin{figure}
     \centering
      \subfloat[Achievable \gls{slnr}]{
      \includegraphics[width=0.472\columnwidth, height =0.18\textwidth]{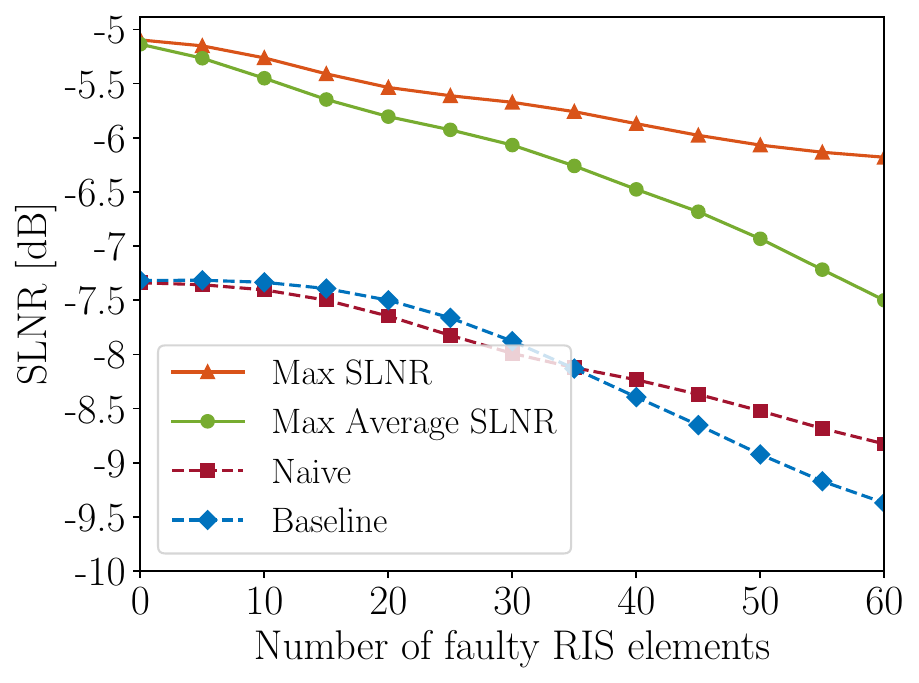}\label{fig:Fig2}
      }
      \hfill
      \centering
      \subfloat[Cost paid in terms of \gls{snr}.]{
      \includegraphics[width=0.472\columnwidth, height =0.18\textwidth]{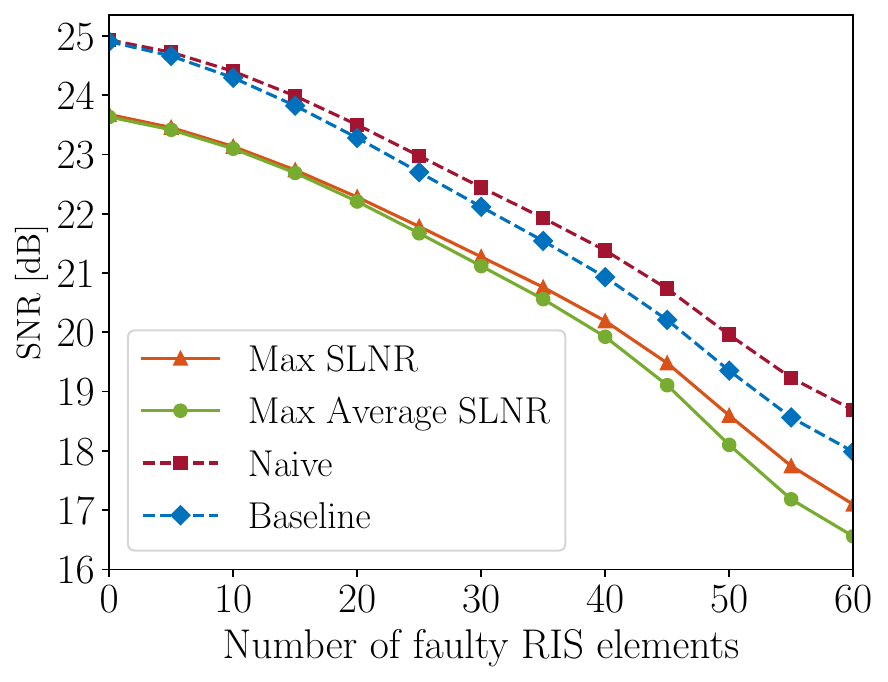}\label{fig:Fig2b}
      }
      \caption{\small Comparison in terms of \gls{slnr} and \gls{snr} between the proposed approaches under perfect and partial \gls{csi}, and the reference schemes, versus the number of faulty \gls{ris} elements.}
      
\end{figure}

%\begin{figure*}[t]
       % \center
        %\includegraphics[width=\textwidth,height=0.6\textwidth]{heatmapplot.pdf}
       % \caption{\small cc }
        %\label{fig:Fig4}
%\end{figure*}

 \begin{figure}[t]
        \centering
        \subfloat[Max SLNR]{\includegraphics[width=0.23\textwidth]{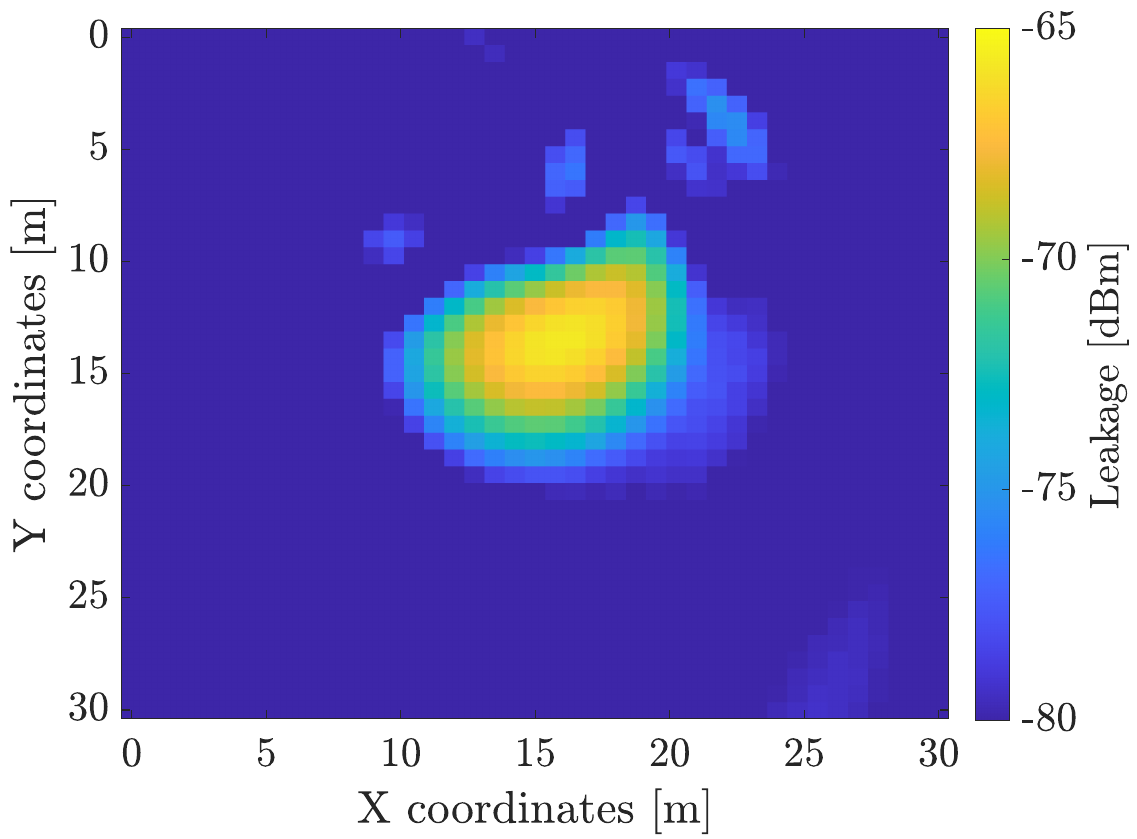}\label{fig:Fig3a}}\quad
        \subfloat[Baseline approach]{\includegraphics[width=0.23\textwidth]{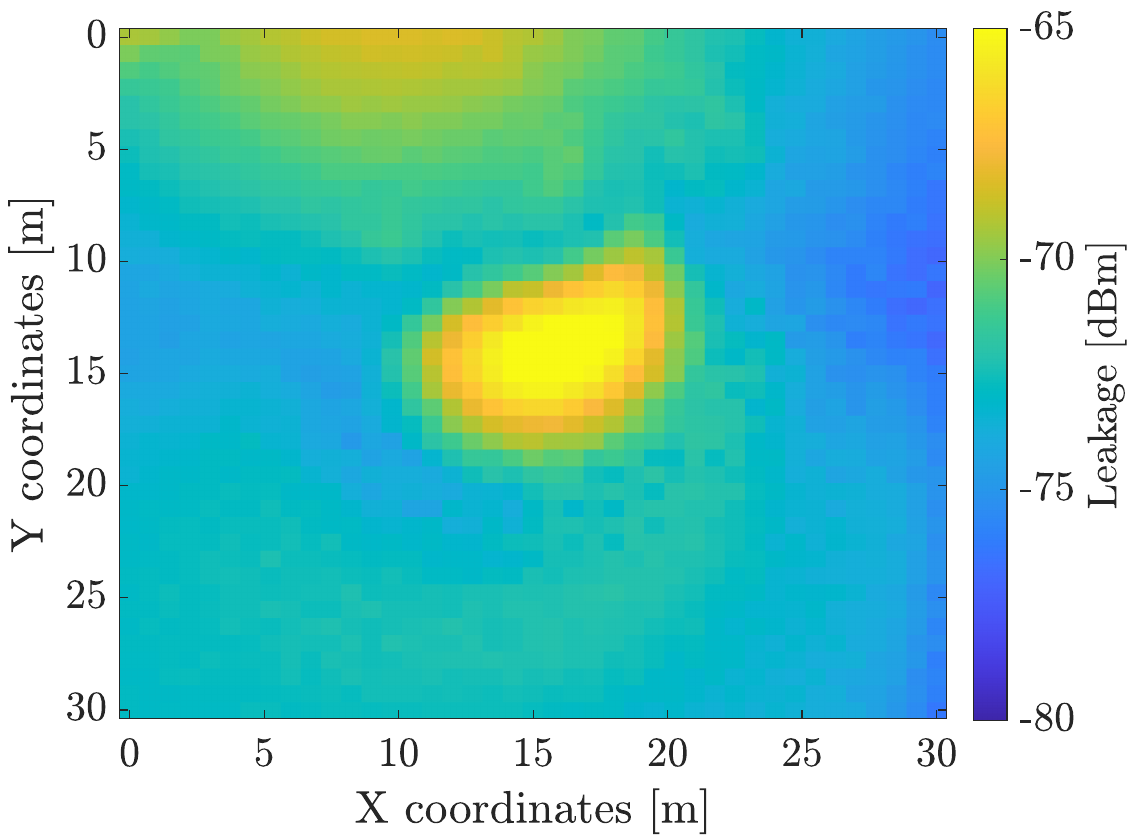}\label{fig:Fig3b}}\\
        \subfloat[Max Average SLNR]{\includegraphics[width=0.23\textwidth]{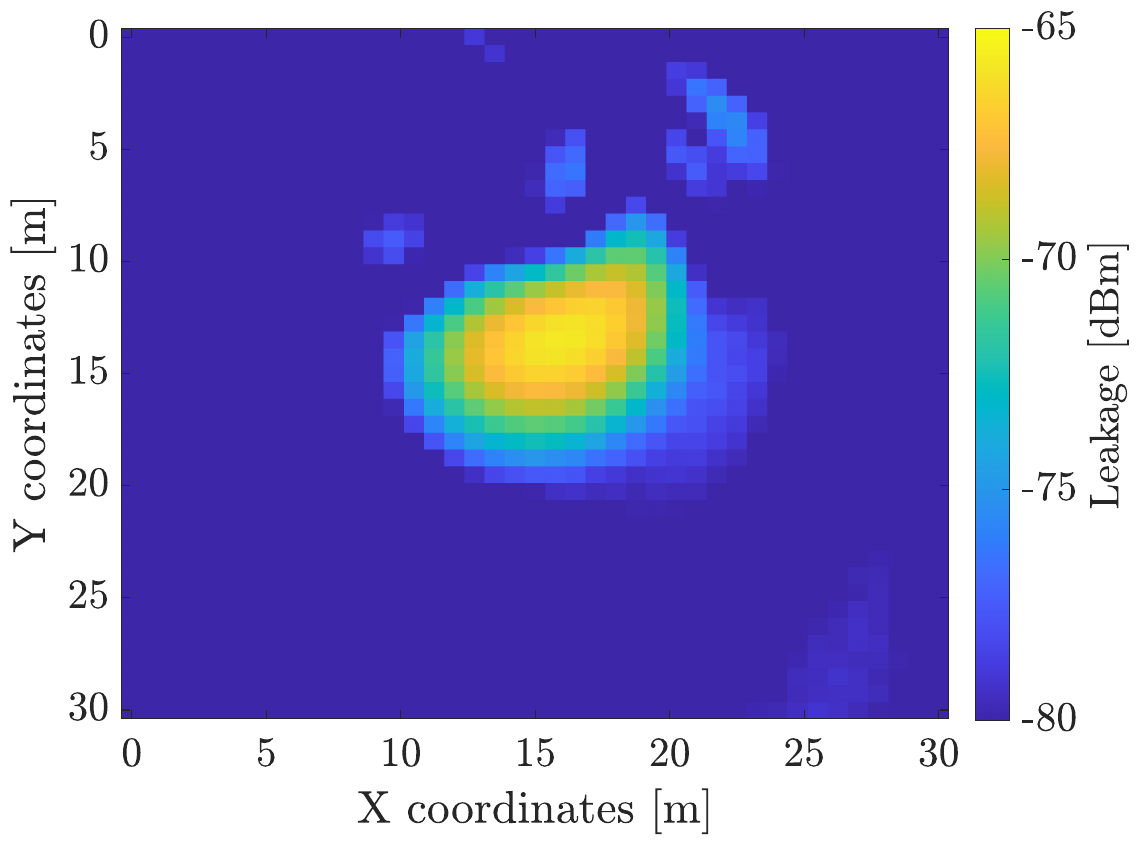}\label{fig:Fig3c}}\quad
        \subfloat[Naive approach]{\includegraphics[width=0.23\textwidth]{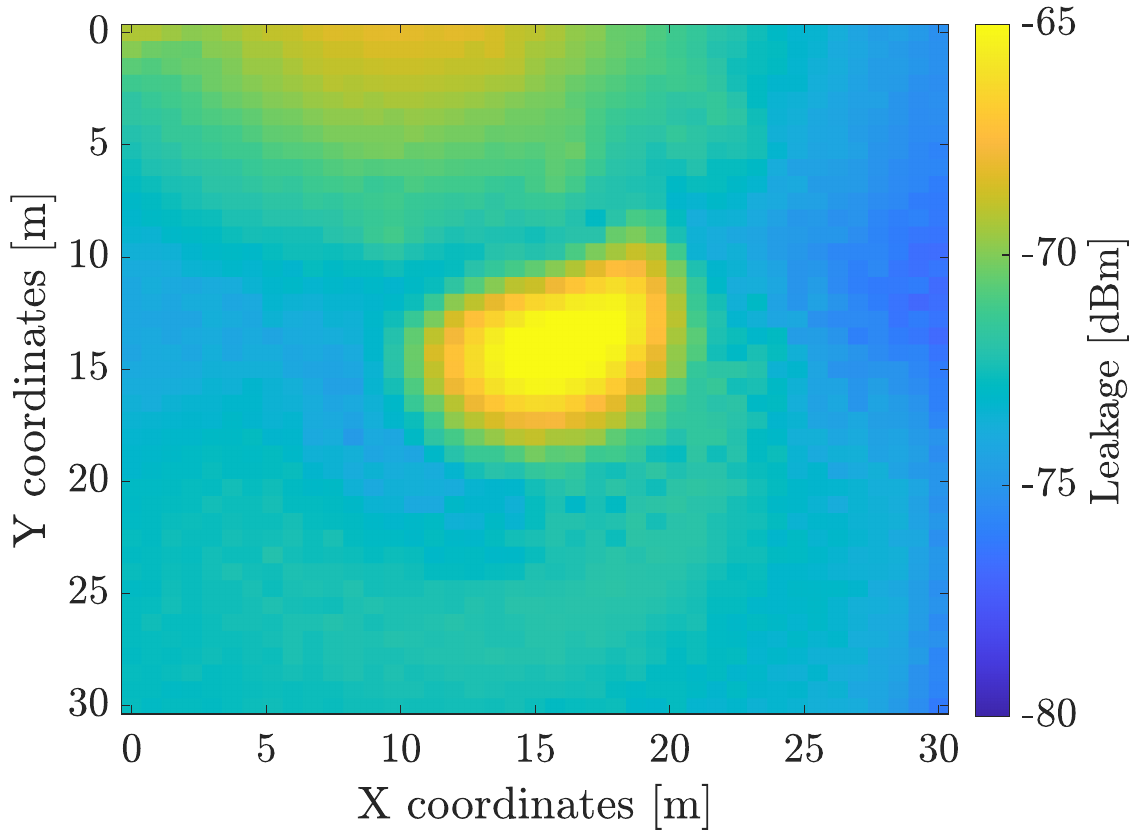}\label{fig:Fig3d}}
        \caption{\small Heatmap of the received power in dBm over the target area with $10$\% faulty \gls{ris} elements for the proposed approaches, i.e., (a) \gls{slnr} maximization under perfect \gls{csi} and (c) \gls{slnr} maximization under partial \gls{csi}, and the reference schemes, i.e., (b) baseline approach and (d) naive approach.}
        \label{fig:Fig3}
         \end{figure}
         
      % \begin{figure*}[t]
      %   \centering
      %   \subfloat[Proposed approach]{\includegraphics[width=0.3\textwidth]{Figures/ht1.pdf}}
      %   \hfill
      %   \subfloat[Baseline approach]{\includegraphics[width=0.3\textwidth]{Figures/ht2.pdf}}
      %   \hfill
      %   \subfloat[Naive approach]{\includegraphics[width=0.3\textwidth]{Figures/ht3.pdf}}
      %   \caption{\small The heatmap of the proposed approach (a), the Baseline approach (b), and the Naive approach (c), for $10$ broken elements in a region with $225$ leakage points located in an area of size $30 \times 30$ in an ordered manner}
      %   \label{fig:Fig4}
      %    \end{figure*}

\begin{comment}
\begin{figure}[t]
        \center
        \includegraphics[width=\columnwidth]{Figures/leakslnr1.pdf}
        \caption{\small \gls{slnr} vs. Leakage point density }
        \label{fig:Fig5}
\end{figure}
\end{comment}
\begin{comment}
 \begin{figure}
     \centering
      \subfloat[\gls{slnr} vs. Leakage point density.]{
      \includegraphics[width=0.47\columnwidth, height=0.3\textwidth]{Figures/leakslnr1.pdf}
      }
      \hfill
      \subfloat[Cost paid for \gls{slnr} maximization vs. leakage point density.]{
      \includegraphics[width=0.47\columnwidth, height=0.3\textwidth]{Figures/leaksnr1.pdf}
      }
      \caption{\small TBD. \textcolor{red}{will fix the ratio of the figures, also the SNR values will hopefully get reduced}}
      \label{fig:Fig6}
\end{figure}   
\end{comment}
\begin{figure}
    \centering
    \includegraphics[width=\columnwidth, height= 0.26\textwidth]{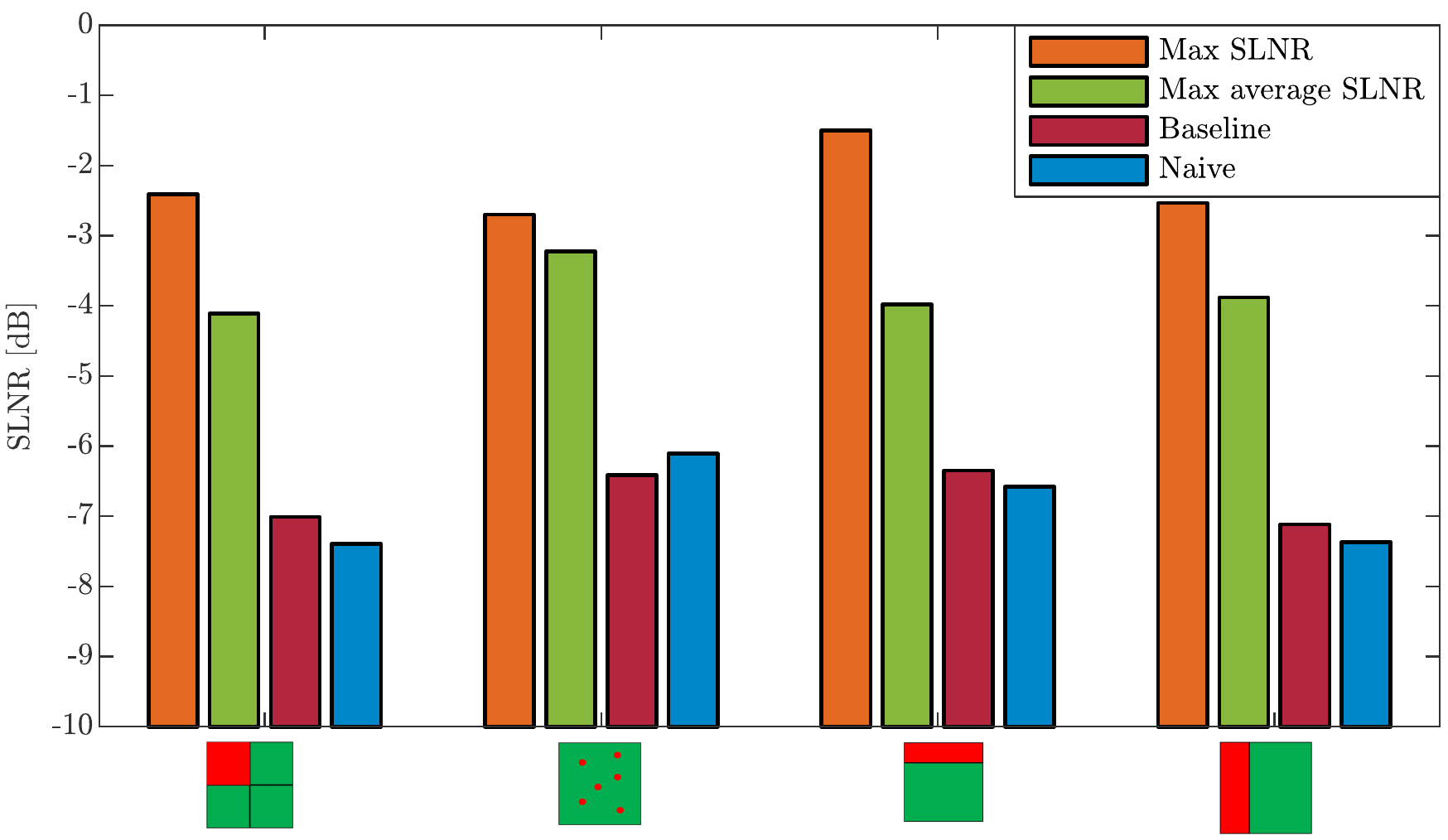}
    \caption{\small \gls{slnr} obtained with the proposed approaches and the reference schemes versus different distributions of $25$\% of faulty antenna elements at the \gls{ris} (marked in red).}
    \label{fig:Fig4}
\end{figure}
%\subsection {Simulation Setup}
%The simulations have been conducted by utilizing an \gls{ris} of size $10 \times 10$ at the carrier frequency of $30$ GHz, with the \gls{ue} and \gls{ris} being located at the two-dimensional coordinates of $(15, 26)$ and $(10, 37)$, respectively. The \gls{ris} is considered to be a \gls{upa} consisting of $N=N_xN_y = 100$ elements with 10 elements on each axis and an inter-element spacing of $d=\frac{\lam\rmBa}{2}$. The \gls{ap} is considered to be transmitting with a power of $P = 8$ dBm per sub-carrier, and $\sigma_{n}^2 = -80$ dBm where  $\sigma_{n}^2$ is the noise power.
%and the computation of the noise power received by the \gls{ue} follows $\sigma_n^2 = WN_0N_f$ where $W = 20$ MHz, $N_0 = -174$ dbm/Hz, and $N_f = 6$ dB are denoted as the signal bandwidth, noise power spectral density, and noise figure respectively \cite{jamali2022}.
%Additionally, for both the \gls{ap}-\gls{ris} and \gls{ris}-\gls{ue} links, we account for $20$ different scattering paths, denoted by $P_G = P_K = 20$. The k-factors and pathloss exponents for the \gls{ap}-\gls{ris} and \gls{ris}-\gls{ue} links. All this info is entered in Table ~\ref{tab:params}. 
%The simulation results are averaged over $1000$ realizations of the channels and the randomly generated values and indices of the broken elements and other system settings, including the coordinates of randomly generated points in the leakage regions.

\subsection{Discussion} 
%{\color{red}\lipsum[1-2]}
%training and testing
In Fig.~\ref{fig:Fig2}, we compare the performance of the proposed approaches under perfect and partial \gls{csi}, denoted as \emph{Max SLNR} and \emph{Max Average SLNR}, respectively, along with the baseline and naive approaches, in terms of the achievable \gls{slnr} versus the number of faulty \gls{ris} elements. For increasing number of faulty \gls{ris} elements, the \gls{slnr} tends to decrease rapidly. However, the proposed approaches obtain significantly higher performance and resilience to faulty \glspl{ris}, especially in the case of perfect \gls{csi} at the cost of a small reduction in \gls{snr}, as shown in Fig.~\ref{fig:Fig2b}. Here, as expected, the naive approach obtains the highest value, though neglecting signal leakage. To better understand the effectiveness of our proposed approaches, in Fig.~\ref{fig:Fig3} we show the \gls{2d} heatmap of the received power over the target area wherein the intended \gls{ue} is located, i.e., the signal leakage, for the case of $10$ faulty \gls{ris} elements. Fig.~\ref{fig:Fig3a} and Fig.~\ref{fig:Fig3c} show considerable improvement in terms of the signal leakage received in the target area as compared to the reference schemes in Fig.~\ref{fig:Fig3b} and Fig.~\ref{fig:Fig3d}. Indeed, for the proposed schemes the signal power is high only in the close-proximity of the \gls{ue}, whereas for the reference schemes the average received power is significantly higher elsewhere, especially close to the \gls{ris} (i.e., the upper-left corner of the figures).

Lastly, in Fig.~\ref{fig:Fig4} we evaluate the impact of the distribution of $25$\% of faulty \gls{ris} elements on the \gls{slnr}. We consider four cases, namely \textit{i)} the upper-left quadrant of the \gls{ris} fails, \textit{ii)} uniform distribution of the faulty \gls{ris} elements (as is the case for the rest of the figures), \textit{iii)} the upper two rows of the \gls{ris} fail, and \textit{iv)} the first two columns starting from the left of the \gls{ris} fail. In all considered scenarios, the proposed schemes outperform the reference schemes. In particular, we infer that having the faulty elements grouped together represents a worst-case scenario for the reference schemes, since this originates strong side lobes pointing towards fixed directions. Conversely, the opposite is true for the proposed schemes: in the case of partial \gls{csi}, the performance is equivalent in all cases, while in the case of perfect \gls{csi} the proposed approach manages to mitigate the unwanted signal spreading especially when the distribution of the faulty elements is not uniform.

\begin{figure}[t]
        \center
        \includegraphics[width=\columnwidth]{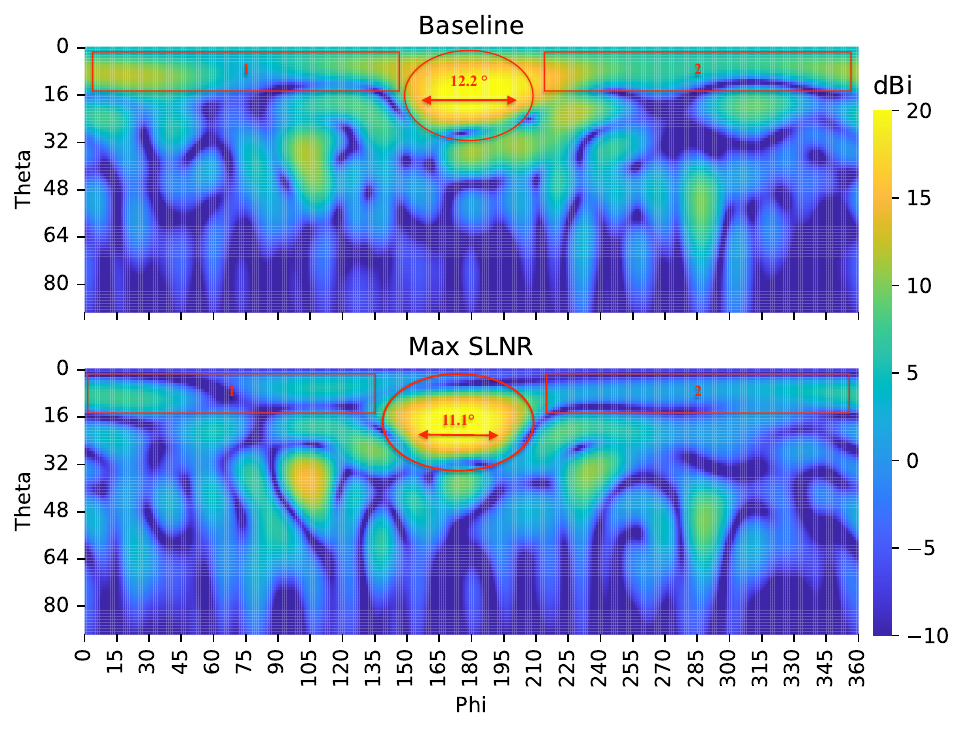}
        \caption{\small Full-wave simulation of the far-field radiation pattern obtained with the proposed approach for perfect \gls{csi} and the baseline.}
        \label{fig:CST}
\end{figure}

%%%%%%%%%%%%%%%%%%%%%%%%%%%%%%%%%%%%%%%%%%%%%%%%%%%%%%%%%%%%%%%%%%
\subsection{Full-wave simulations}\label{sec:CST}
\begin{figure}[t]
        \center
        \includegraphics[width=\columnwidth]{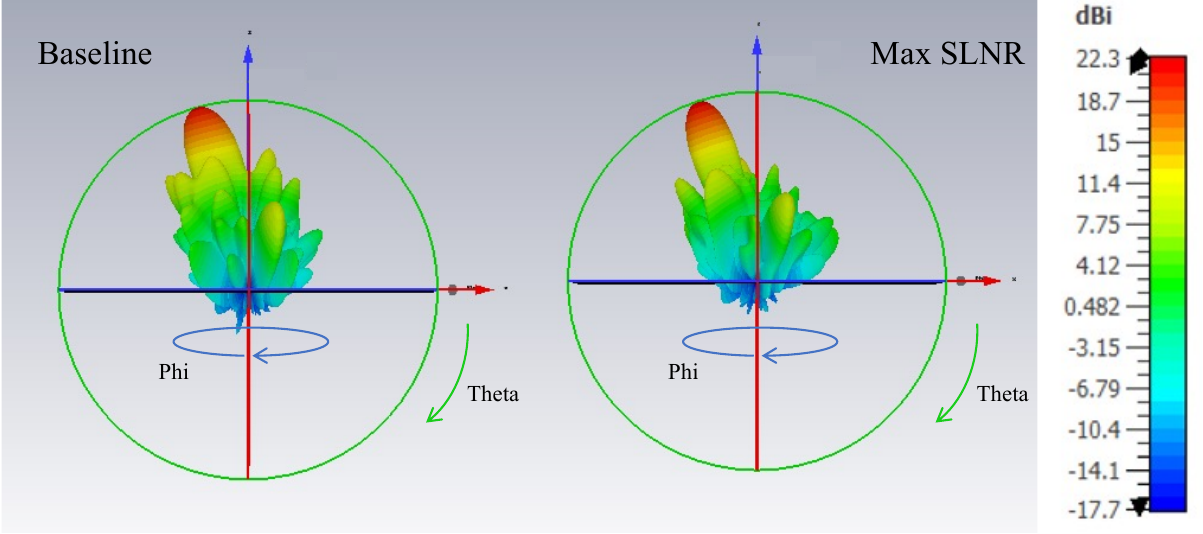}
        \caption{\small Full-wave simulation of the $3$D beampattern obtained with the proposed approach for perfect \gls{csi} and the baseline.}
        \label{fig:Fig6}
\end{figure}

%%%%%%%%%%%%%%%%%%%%%%%%%%%%%%%%%%%%%%%%%%%%%%%%%%%%%%%%%%%%%%%%%%
In this section, we demonstrate the effectiveness of our proposed approach in the case of perfect \gls{csi} versus the baseline scheme for the case of $10$\% faulty \gls{ris} elements using a full-wave simulator, CST Studio Suite 2019. Fig.~\ref{fig:CST} shows the far-field radiation pattern as the spherical coordinates \emph{Theta} and \emph{Phi} vary along the azimuth and elevation directions, respectively. In the range of $[0^\circ, 15^\circ]$ in azimuth, the baseline approach exhibits an average power of $16.8$ and $16.3$~dBi, which are marked with two red boxes, respectively, and a main beam with a half power beamwidth of $12.2^\circ$. On the contrary, the proposed approach manages to significantly reduce the power outside the main beam pointing towards the \gls{ue} (average of $11.2$ and $9.8$~dBi, respectively), with its half power beamwidth reduced to $11.1^\circ$. Hence, the proposed approach mitigates the undesired side lobes while better focusing the power towards a specific direction. Indeed, when there are faulty elements on the \gls{ris}, the behavior of the \gls{ris} becomes similar to an omnidirectional antenna, thus reflecting the signal to all directions. This effect can be clearly seen in Fig.~\ref{fig:Fig6}, which shows the full-wave simulation of the $3$D beampattern for the considered schemes. As expected, the proposed approach exhibits a narrower main beam and lower side lobes as compared to the baseline scheme.

%%%%%%%%%%%%%%%%%%%%%%%%%%%%%%%%%%%%%%%%%%%%%%%%%%%%%%%%%%%%%%%%%%
%{\color{red}\lipsum[1]}

%\begin{figure}[t]
        %\center
        %\includegraphics[width=\columnwidth]{Figures/leaksnr1.pdf}
        %\caption{\small  Cost paid for \gls{slnr} maximization vs. leakage point density}
        %\label{fig:Fig6}
%\end{figure}     

%%%%%%%%%%%%%%%%%%%%%%%%%%%%%%%%%%%%%%%%%%%%%%%%%%%%%%%%%%%%%%%%%%
\section{Conclusions}\label{sec:Conclusions}
%%%%%%%%%%%%%%%%%%%%%%%%%%%%%%%%%%%%%%%%%%%%%%%%%%%%%%%%%%%%%%%%%%
In this paper, we have proposed a novel mathematical framework to address the problem of mitigating unintentional signal leakage caused by faulty \gls{ris} elements, which may result in additional interference to non-intended \glspl{ue} and security threats to potential eavesdroppers. Specifically, we have formulated a low-complexity model for the random signal attenuation and phase shift introduced by each faulty \gls{ris} element, and we have proposed a proven convergent iterative algorithm that targets the optimization of the functioning \gls{ris} elements by maximizing the \gls{slnr} in a given \gls{2d} area wherein the intended \gls{ue} is located. This approach is applicable in the case of both perfect \gls{csi}, i.e., when the location and state of the faulty elements are known, and in the case of partial \gls{csi}, i.e., when the state of the faulty elements is unknown. Numerical results have demonstrated the effectiveness of the proposed approaches in terms of achievable \gls{slnr} (up to $35$\% improvement), at the cost of a small reduction in \gls{snr} (within $4$\%) as compared to other reference schemes. 

Several intriguing research directions, involve the exploration of network challenges in cases of multiple \glspl{ris} or multiple \glspl{ue}, and refining optimization problems related to these contexts.

\bibliographystyle{IEEEtran}
\bibliography{refs}
\end{document}